\documentclass[prl,twocolumn]{revtex4}

\usepackage{graphicx}
\usepackage{dcolumn}
\usepackage{bm}

\begin{document}

\preprint{APS/123-QED}

\title{Simulation of quantum dead-layers in ferroelectric tunnel junctions}

\author{K. M. Indlekofer}
\email{m.indlekofer@fz-juelich.de}
\affiliation{
Institute for Thin Films and Interfaces (ISG-1) and
Center of Nanoelectronic Systems for Information Technology (CNI),
Research Centre J\"ulich GmbH, D-52425 J\"ulich, Germany
}
\author{H. Kohlstedt}
\affiliation{
Institute for Solid State Research (IFF) and
Center of Nanoelectronic Systems for Information Technology (CNI),
Research Centre J\"ulich GmbH, D-52425 J\"ulich, Germany
}

\date{\today}

\begin{abstract}
In this letter, we simulate electronic transport through
a metal-ferroelectric-metal tunnel junction by use of a nonequilibrium
Green's function approach.
We show that quantum effects such as Friedel oscillations
lead to deviations from the Thomas-Fermi screening model.
As a consequence, we predict a bistable resistive switching effect,
depending on the polarization state of the ferroelectric tunnel barrier.
\end{abstract}

\pacs{XXX}
\maketitle

The finite penetration depth of the electric field in a metallic system
leads to considerable corrections to the electronic properties
of a metal-insulator-metal tunneling structure \cite{kie96}.
For example, Mead \cite{mea61} observed a deviation from the classical
geometric capacitance arising from the
electric field penetration into the metal electrodes.
This finding was explained theoretically by Ku and Ullman \cite{ku64},
as well as Simmons \cite{sim65},
by modeling the total capacitance as a sum of
a (geometric) dielectric capacitance and electrode capacitance contributions
(due to the finite field penetration).
Black and Welser extended this approach
to high-k materials (for example BaSr$_x$Ti$_{1-x}$O$_3$) \cite{bla99}.
In general, the field penetration effect is pronounced in capacitors
and tunnel junctions containing
ultrathin insulator films with a high dielectric permittivity.
In addition, field penetration was also noticed in the context of finite
size effects of complex oxide ferroelectric
capacitors \cite{meh73,jun03,daw02,daw03}.
The finite screening length furthermore leads to uncompensated ferroelectric
charge layers at metal-ferroelectric interfaces under short circuit
conditions.
First-principles calculations by Junquera and Ghosez 
predicted ferroelectric instabilities in SrRuO$_3$/BaTiO$_3$/SrRuO$_3$
heterostructures with a tunnel barrier thickness even down to $2.4$nm.
(In fact, recent experiments on Lead compound oxides demonstrated
ferroelectricity in ultrathin films \cite{tyb99,fon04}).
In all examples, the field penetration in the metal electrodes was described on
the basis of the Thomas-Fermi screening length \cite{kie96}.

\begin{figure}
\includegraphics[width=7cm]{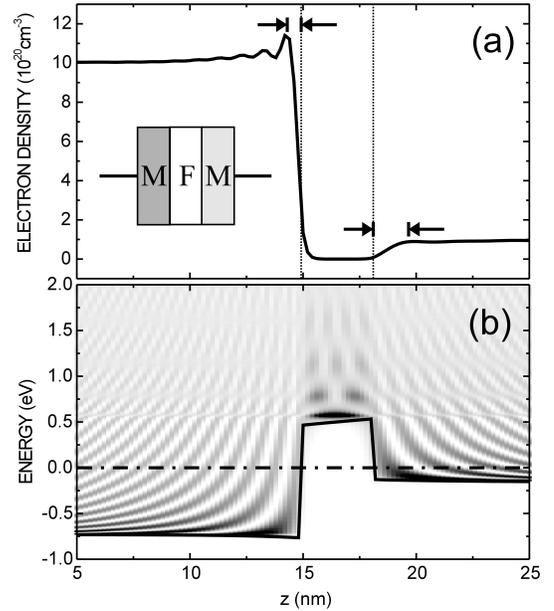}
\caption{\label{fig:fig1}
Simulated electron density distribution (a) and potential profile (b)
(with LDOS grayscale plot) for zero bias.
(The Fermi energy corresponds to the dash-dotted line.)
The inset shows a schematic sketch of the tunnel junction.
}
\end{figure}

In this letter, we show that quantum interference
effects such as Friedel oscillations
\cite{kie96}
lead to deviations from the semiclassical Fermi-Thomas screening
within a metal-insulator-metal tunnel junction.
Recently, we presented the concept of a ferroelectric tunnel junction (FTJ)
\cite{koh02,rod03} based on a metal-ferroelectric-metal (MFM) layer sequence,
consisting of two (semi-) metals separated by an
ultrathin ferroelectric layer as visualized in the inset of
Fig.~\ref{fig:fig1}(a).

In order to simulate the IV characteristics of the FTJ,
a non-equilibrium Green's function approach has been
used \cite{wingreen,ind02,schaefer}.
The employed empirical tight-binding one-band formulation constitutes
a good approximation for the description of conduction band electrons inside
the ferroelectric tunnel barrier and the (semi-) metallic electrodes.
Within this framework, Coulomb charging effects are accounted for in terms
of a self-consistent Hartree-potential.
Polarization charges at the two interfaces of the ferroelectric material
are implemented as charged mono-layers.
(For simplicity, we assume a mono-domain polarization state with a
polarization vector perpendicular to the junction plane.
The device structure is translationally invariant within the grown layers.)
An applied bias voltage is implemented as a difference in the local
Fermi-energies of the two reservoirs.

In the following, we will discuss an idealized case of an MFM structure,
assuming an effective electron mass of $m^*=0.5 m_e$
and a dielectric constant of $\epsilon_r=300$ throughout the device.
The ferroelectric tunnel barrier of $3.2$nm and $0.5$eV height
(with interface polarization charges of $\sigma=\pm 5\times 10^{13}$cm$^{-2}$)
is surrounded by two (semi-) metallic
contact regions with electron concentrations of $10^{21}$cm$^{-3}$ and
$10^{20}$cm$^{-3}$, respectively.
(Of course, the chosen parameters have to be adjusted to the individual
experimental device structure.)

Fig.~\ref{fig:fig1}(a) shows the simulated electron contration under
zero bias conditions (short circuit case).
Due to the wave-nature of the electron, Friedel oscillations arise
within the electrode regions in the vicinity of the tunnel barrier
(clearly pronounced in the left contact).
The oscillation period is given by the Fermi wavelength divided by two
($\lambda_F/2\approx 1.0$ nm in the l.h.s. reservoir) \cite{kie96}.
As an important consequence, quantum mechanically determined
partially depleted regions close to the tunnel barrier arise
(see arrows in Fig.~\ref{fig:fig1}(a): $1.0$nm and $2.2$nm),
in contrast to the abrupt electron concentration drop
in the classical case.
We refer to this phenomenon as a "quantum dead-layer"
(analogous to a conventional dead-layer \cite{dro59}
which also introduces a non-screening and non-ferroelectric
region at the metal-insulator interface).
Within the semi-metallic electrode
on the r.h.s., the depleted region is further extended due to band-bending,
varying with the applied voltage.
In Fig.~\ref{fig:fig1}(b), the corresponding self-consistent potential profile
is shown (solid line) with a grayscale plot for the local density of states
(LDOS).
One can clearly identify the polarization charge at the interfaces of the
ferroelectric tunnel barrier in terms of a kink in the continuous part of
the potential and a tilted barrier potential due to the resulting
depolarization field.
The LDOS plot reveals the wave-properties of the electron:
Firstly, standing wave patterns result from reflections at the tunnel barrier,
giving rise to the Friedel oscillations with a quantum dead-layer.
Secondly, multiple reflections within the barrier region yield resonances in
the higher energy region.
(Latter may become relevant for hot-carrier injection conditions.)
Furthermore, within the reservoir on the l.h.s., a shallow triangular
quantum well arises, which is responsible the screening charge.

\begin{figure}
\includegraphics[width=7cm]{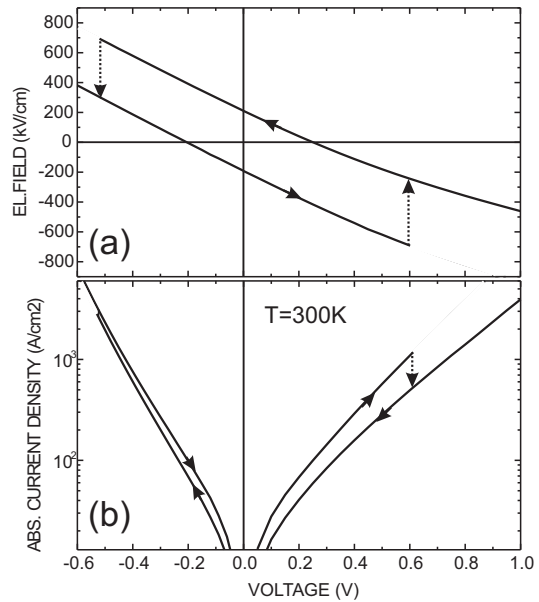}
\caption{\label{fig:fig2}
Simulated electric field (a) and current-density (b) characteristics.
(The assumed intrinsic coercitive field is $700$kV/cm.)
}
\end{figure}

The local electric field within the ferroelectric barrier as a
function of the applied voltage is shown in Fig.~\ref{fig:fig2}(a).
Here, an intrinsic coercitive field of $700$kV/cm has been assumed,
resulting in transition voltages of approximately $+0.6$V and $-0.5$V.
The asymmetry stems from the assumed electron concentration asymmetry within
the electrodes.
Since there is a significant voltage drop within the electrode
regions close to the barrier (depleted regions),
the external voltage necessary to induce a polarization flip in the
ferroelectric material is significantly higher than the intrinsic voltage drop
across the barrier.
Additionally, in Fig.~\ref{fig:fig2}(b), the corresponding
current-density vs. voltage characteristics is plotted in a log-scale,
revealinge an exponential behavior of the current density
which is determined by the tunnel barrier.
Depending on the the polarization state and the sign of the applied voltage,
the slope in the log-scale
varies due to different potential profiles.
The bistable IV characteristics could make such an MFM 
device a possible candidate for a non-volatile memory element.
Obviously, the total capacity of the considered device structure will also be
affected by the quantum dead-layer and depletion effects discussed above.

In conclusion,
we have considered quantum interference effects
in ultrathin metal-ferroelectric-metal tunnel junctions.
The current-voltage charateristics of an exemplary tunnel structure
has been simulated with the help of
a nonequilibrium Green's function formalism.
As a result, a bistable resistive switching behavior was predicted.

\begin{acknowledgements}
This work was supported by the the
Volkswagen-Stiftung Project Nano-sized ferroelectric hybrids under
Contract No. I/77737.
The financial support by the Deutsche Forschungsgemeinschaft (DFG)
is gratefully acknowledged.
\end{acknowledgements}

\bibliography{paper}

\end{document}